\documentclass[sort&compress]{elsarticle}

\usepackage{lineno,hyperref}
\modulolinenumbers[5]

\usepackage{lscape} 
\usepackage[usenames]{color} 

\journal{Chemical Physics Letters}

\begin{document}

\begin{frontmatter}

\title{Shannon entropy: a study of confined hydrogenic-like atoms}

\author{Wallas S. Nascimento\fnref{1}}
\author{Frederico V. Prudente\fnref{2}}
\address{Instituto de F\'isica, Universidade Federal da Bahia, 40170-115 Salvador, BA, Brasil}
\fntext[1]{wallassantos@gmail.com}
\fntext[2]{prudente@ufba.br}




\begin{abstract}
The Shannon entropy in the atomic, molecular and chemical physics context is presented by using as test cases the hydrogenic-like atoms H$_c$, He$_c^{+}$ and Li$_c^{2+}$ confined by an impenetrable spherical box. Novel expressions for entropic uncertainty relation and Shannon entropies $S_r$ and $S_p$ are proposed to ensure their physical dimensionless characteristic. The electronic ground state energy and the quantities $S_r$, $S_p$ and $S_t$ are calculated for the hydrogenic-like atoms to different confinement radii by using a variational method. The global behavior of these quantities and different conjectures are analyzed. The results are compared, when available, with those previously published.
\end{abstract}


\end{frontmatter}


\section{Introduction}

The basis of the mathematical theory of communication or information theory was established by Shannon in 1948~\cite{shannonoriginal1}, where it was analyzed substantially the communication process, i.e., how is the interaction between different mechanisms. This connection between mechanisms, usually by sending and receiving a message (oral or written forms, pictures, music, etc.), happened to have a statistical and probabilistic approach, making their semantic aspects irrelevant with regard to the engineering ones. Information in this field is not the meaning, but it is a measure of freedom to select a message among many others~\cite{livro_teoria_matemcatica_da_comunicacao_original}. A fundamental concept is the information entropy or Shannon entropy which has complementary interpretations - quantity of information (after measurement) or uncertainty (before measurement) - in a given probability distribution~\cite{nielsen}.

Information theory, through of the Shannon entropies in the position ($S_r$) and momentum ($S_p$) spaces, has been applied in the context of the quantum theory from the 1980s~\cite{livro_sen_shannon,tunelamento_shannon,information_thermodynamics}. More specifically, we can cite the derivation of analytical relations in systems such as a particle confined in a box~\cite{relacoessomaentropica2,particula_caixa_shannon2}, harmonic oscillator~\cite{relacao_de_incerteza_oscilador,shannonconfinadohosciladortempo} and hydrogen atom~\cite{relacao_analitica_oscilador_hidrogenio}. Moreover, the treatment of electronic excitations~\cite{excitacao_shannon,excitacao_shannon_confinado2} and complexity of systems \cite{complexidade1,complexidade_3} and characterization of chemical processes \cite{shannon_reacao_quimica,shannon_reacao_quimica_2} have also used the Shannon entropy in their analysis. And in recent years the Shannon entropy has been studied from the perspective of spatially confined quantum systems~\cite{entropiaatomosconfinados,aquino,shannonconfinado2,entropy_for_an_infinite_circular_well,benchmark_hidrogenio, oscilador_confinado_simetrico_assimetrico}.

Confined systems have their physical and chemical properties modified in relation for the free system. In a first approximation, atoms and molecules under high pressure, quantum dots and dense astrophysical objects can be treated as confined systems in spherical barrier of infinite potential (see Refs.\cite{confinadosistemacorte,marcilioartigoconfinado} and references therein). Thus, such systems have been the subject of numerous studies with different methods to impose the confinement, making this a relevant field of research~\cite{jaskolski,marcos,livro_sen_confinado,livro_sistemas_confinados1,prudente,fred_ginette}. A particularly interesting situation is when the confinement is rigorous (strong confinement regime). For example, it was shown that in such regime the energies of confined hydrogenic-like atoms tends to the values of a particle confined in a spherical cage~\cite{barbosa}. Recent studies for the harmonic potential showed that such analysis can be made based on Shannon entropy~\cite{wallas_fred_educacional,uncertainties_of_the_confined_harmonic}. 

The entropy sum ($S_t$), quantity defined as the sum of the Shannon entropies in the position and momentum spaces, has occupied a privileged place in the study of quantum systems in the information theory context. For instance, an entropic uncertainty relation that it has been treated as a stronger version of the Heisenberg uncertainty relation can be derived from the entropy sum~\cite{relacaodeincertezainformacao}. Additionally, we can cite the conjectures pointing to possible universal relationships of that quantity~\cite{coparacaofermionboson}, as well as the promising study of electron correlation based on the value of the entropy sum~\cite{shannon_correlacao3}. More specifically, studies have shown that improvements in basis function set (according to variational energy criteria) leads to an increase of the entropy sum for atomic systems. This leads to the conjecture that $S_t$ can be regarded as an indicator of the quality of basis functions~\cite{gadrequalidade,gardre1985,estudo_numerico_qualidade_base}, proposal known as Gadre-Sears-Chakravorty-Bendale (GSCB) conjecture. Furthermore, previous works show that the value of the entropy sum not depend of the atomic number for fundamental or excited state of free hydrogenic-like atoms, while in an isoelectronic multielectronic atoms series its value depends of the atomic number~\cite{shannon_correlacao}.   

Despite the success of using the Shannon entropy in atomic, molecular and chemical physics, one aspect that still requires discussion is its dimensional problem. Originally, the informational entropy proposed by Shannon is defined in terms of a logarithmic function from a discrete or continuous dimensionless probability distribution, being expressed in units of information bits, hartleys, nats and etc, according to the base of logarithm employed (a base change results only in a scale change). In this sense Shannon's logarithmic quantities acquire information-theoretic units. For the other hand, in the quantum theory the probability density functions in the position and momentum spaces have dimensional characteristics. In this framework the use of the quantum probability density functions to build the Shannon entropy requires adaptations in the expression originally proposed because of their physical dimensionality.  

Recent works have already proposed redefinitions of the Shannon entropy in the position spaces to solve this dubiousness explicitly. For example it is inserted into the logarithmic argument a function having the dimensions of reciprocal volume (expressed in the appropriate units) \cite{teoriadainformacaodesistemasquanticos} or of electronic densities measured on the nucleus of the atom~\cite{aproximacao_dimencao_shannon}.  Note that these proposals were presented with specific objectives, the first one to study problems with a delimited contour of the electronic probability density, while the second one to define an expression for the Shannon entropy non-negative under all space.

In the present work, we have developed a theoretical study on one-electron atomic systems confined within impenetrable walls, more precisely the hydrogen (H$_c$), helium cation (He$_c^{+}$) and lithium dication (Li$_c^{2+}$), by using the Shannon entropies and entropy sum. Initially, we suggest modified and more general expressions to the entropy sum (and the correspondent entropic uncertainty relation) and Shannon entropies $S_r$ and $S_p$ that ensure their physical dimensionless characteristic. Subsequently, we have performed the calculations of the energy, $S_r$, $S_p$ and $S_t$ of H$_c$, He$_c^{+}$ and Li$_c^{2+}$ in the ground state for different confinement radii by using a variational method. The global behavior of these quantities are analyzed and two conjectures are tested, the first on the dependence of them due the atomic number and the second on the minimum value of $S_t$ with respect to the confinement radius. Additionally, we have analyzed the entropic uncertainty relation validity and the use of $S_t$ as a measure of basis function set quality, GSCB conjecture, in confined environments (for neutral and ionized atoms). And, finally, we have examined the strong confinement regime. 

The paper is organized as follows: in section~\ref{sec:2} are presented the concepts and definitions of the information theory in the quantum mechanics context and the novel expressions of $S_t$, $S_r$ and $S_p$. In section~\ref{sec:3} the systems of interest are presented and the trial ground state wave functions are proposed. The results for the confined hydrogenic-like atoms are shown in section~\ref{sec:4}, when pertinent analysis are done. And, finally, in the last section the main aspects of the present work are summarized and the concluding remarks are presented.

\section{Concepts}\label{sec:2}

In the context of the quantum theory is known that the measure of any two observables $\cal{A}$ and $\cal{B}$ can only be done within of a limit of accuracy, resulting at the standard uncertainty relation. In particular, in the case of the position and momentum observables  we have the Heisenberg uncertainty principle. On the union of the quantum mechanics and information theory is obtained the entropic uncertainty relation, being treated as a stronger version of the Heisenberg uncertainty relation. 

Usually the entropic uncertainty relation is derived of the entropy sum, a quantity defined as $S_t=S_r+S_p$, being $S_r$ and $S_p$ the Shannon entropies in the position and momentum spaces, respectively. The $S_r$ and $S_p$ quantities are defined by replacing the dimensionless probability distribution in the original Shannon entropy formulation with the quantum probability densities in the position and momentum spaces. However, in quantum theory the probability densities $\rho(\vec{r})$ and $\gamma(\vec{p})$, respectively, in the $n$-dimensional  position and momentum spaces are normalized to unity and have, respectively, dimension of inverse of  $n$-dimensional position and momentum volume. More precisely, $\rho(\vec{r})$ and $\gamma(\vec{p})$ are given, respectively, in terms of ${\arrowvert \psi (\vec{r})\arrowvert}^2$ and ${\arrowvert \widetilde{\psi} (\vec{p})\arrowvert}^2$, being $\psi (\vec{r})$ and $\widetilde{\psi} (\vec{p})$ the wave functions in position and momentum spaces.

Thus, to avoid the dimensionality problem of $\rho(\vec{r})$ and $\gamma(\vec{p})$, we propose the following form for the entropic uncertainty relation in the context of the atoms and molecules study:
\begin{equation}
S_t = - \int \int \ d\vec{r} \ d\vec{p} \ \rho(\vec{r}) \ \gamma(\vec{p}) \ \ln( \ {\hbar}^n \ \rho(\vec{r}) \ \gamma(\vec{p})) \geq n(1+\ln\pi) \ ,
\label{soma_entropica}
\end{equation}
where $n$ is the dimension on the position (and the momentum) space. The relation (\ref{soma_entropica}) reaches equality when it is used a Gaussian wave functions. Moreover, this relation has the property of being invariant when subjected to a scale transformation, in other words, is unaffected by a uniform elongation or compression of the atom~\cite{gardre1985}. 

Note that the relation (\ref{soma_entropica}) can be decomposed in the sum of the $S_r$ and $S_p$. Defining ${\hbar}^n = \lambda . \kappa $, we can write,
\begin{equation}
S_t = -\int \int \ d\vec{r} \ d\vec{p} \ \rho(\vec{r}) \ \gamma(\vec{p}) \ \ln( \  \lambda . \kappa \ \rho(\vec{r}) \gamma(\vec{p})) \ . 
\end{equation}
Using the properties of the logarithmic function we have
\begin{eqnarray}
S_t &=& -\int \int \ d\vec{r} \ d\vec{p} \ \rho(\vec{r}) \ \gamma(\vec{p}) \ [\ln( \  \lambda \rho(\vec{r})) + \ln(\ \kappa \gamma(\vec{p}) )] \nonumber \\ 
    &=& -\int \ d\vec{r} \ \rho(\vec{r})\ln(\lambda \ \rho(\vec{r}))-\int \ d\vec{p} \ \gamma(\vec{p})\ln(\kappa \ \gamma(\vec{p})) \ .
\end{eqnarray}
Thus, we can define
\begin{equation}
S_r=- \int \ d\vec{r} \ \rho(\vec{r})\ln(\lambda \ \rho(\vec{r}))
\label{entropia_posicao}
\end{equation}
and
\begin{equation}
S_p=-\int \ d\vec{p} \ \gamma(\vec{p})\ln(\kappa \ \gamma(\vec{p})) \ ,
\label{entropia_momento}
\end{equation}
where the quantities $S_r$ and $S_p$ are, respectively, the Shannon entropies in the position and momentum spaces.

The $\lambda$ and $\kappa$ constants ensure that the arguments of the logarithm function are dimensionless. An interpretation for $S_r$ is to be a measure of the uncertainty in the location of the particle position, while $S_p$ can be considered as a measure of the uncertainty in the determining of the particle momentum. Still, the Shannon entropy is a measure more satisfactory of the uncertainty (hence of the spread of the probability distribution) when compared to the standard deviation~\cite{quantuminformacao,teoriadainformacaoesqueezingdeflutuacoes}.

Note that the informational entropy introduced by Shannon is based on discrete or continuous probability distribution and has the characteristic of being dimensionless in the physical point of view. The Shannon $S_r$ and $S_p$ entropies [Eqs.~(\ref{entropia_posicao}) and (\ref{entropia_momento})] proposed by us differ from those previously reported in the literature by the introduction of the constants $\lambda$ and $\kappa$, respectively, with dimensions of position and momentum volume. The natural choice for these constants in the scope of the quantum theory is the following: $\lambda = {a_0}^n$ and  $\kappa = {(\frac{\hbar}{a_0})}^n$, being $a_0$ the Bohr radius. We point out that the proposed forms are independent of the units system used and of the quantum system to be treated (including systems of arbitrary dimensions, see Ref. \cite{aquilanti1997}). In this sense they are more general in relation to the usual expressions. Furthermore, by using atomic units ($a_0$ and $\hbar$ equal to 1 atomic unit), the relation (\ref{soma_entropica})  and the Eqs.~(\ref{entropia_posicao}) and (\ref{entropia_momento}) assume the current forms of the literature, however, now with a consistent dimensional formulation.

\section{Methodology}\label{sec:3}

The time independent radial Schr\"odinger equation for confined hydrogenic-like atoms in atomic units, assuming an infinite mass for the nucleus and putting it on the center of the hard sphere with radius $r_c$, is given by 
\begin{equation}
[-\frac{1}{2 r} \frac{{\rm d}^2}{{\rm d}r^2} r+\frac{l(l+1)}{2 r^2} +V(r)]\psi (r)=E \psi(r) \ ,
\label{eqesfericas}
\end{equation}
with 
\begin{equation}
V(r)=  \left \{ \begin{array}{ccc}
-\frac{Z}{r} &  & r < r_c \\
\infty &  & r \geq r_c   
\end{array}\right.  \ ,
\label{potencialcomconfinado}
\end{equation}
being $\psi (r)$ the radial wave function solution, $l$ the angular momentum quantum number, $E$ the energy for the stationary state and $Z$ the atomic number ($Z=1$, $Z=2$ and $Z=3$, respectively, for the H$_c$,  He$_c^{+}$ e  Li$_c^{2+}$ confined atoms). 

The variational method~\cite{sakurai} is employed to obtain the ($l=0$) ground state solution of Eq.~(\ref{eqesfericas}). For this, we have employed a trial wave function of the type: 
\begin{equation}
\tilde{\psi}(r)= \phi^c(r)= Ae^{-\alpha Z{r}}\Omega^c(r) \ ,
\label{solfudamentalconfinado}
\end{equation} 
where $\Omega^c(r)$ is a cut-off function that satisfies $\Omega^c(r_c)=0$ condition, $A$ is a normalization constant and $\alpha$ is the parameter to be determined minimizing the total energy functional. 

In the literature there are several proposals for the cut-off function~\cite{ludena,varshni} and we adopted three possible ones. The first one is the polynomial function
\begin{equation}
\Omega_{1q}^c (r)=[1-(\frac{r}{r_c})^q] \ \  \Longrightarrow \ \ \phi_{1q}^c(r)
\label{funcaodecortetipodois}
\end{equation}
where $q$ (= 1, 2 and 3) defines the polynomial degree. The other two types are the trigonometric functions as follows:
\begin{equation}
\Omega_2^c (r)=[\sin (1-\frac{r}{r_c})] \ \  \Longrightarrow \ \ \phi_{2}^c(r)
\label{funcaodecorte2}
\end{equation}
and
\begin{equation}
\Omega_3^c (r)=[\cos (\frac{\pi}{2}\frac{r}{r_c})] \ \  \Longrightarrow \ \  \phi_{3}^c(r). 
\label{funcaodecorte3}
\end{equation}
Thus, the trial wave functions proposed here to describe the confined hydrogenic-like atoms are in the form of the notation and with the appropriate labels given by $\phi_{11}^c(r)$, $\phi_{12}^c(r)$ , $\phi_{13}^c(r)$, $\phi_{2}^c(r)$ and $\phi_3^c(r)$. 

\section{Results and discussion}\label{sec:4}

The ground state wave functions and energies of confined H$_c$, He$_c^{+}$ e Li$_c^{2+}$ were determined for each type of trial wave functions and different values of $r_c$ by the choice of the optimal variational parameter $\alpha$. From the knowledge of $\tilde{\psi}(r)$, we were able to determine Shannon entropies $S_r$ and $S_p$ [Eqs.~(\ref{entropia_posicao}) and~(\ref{entropia_momento})] and the entropy sum $S_t$ [Eq.~(\ref{soma_entropica})]. The results obtained are summarized in Tables~\ref{energias_soma_entropica_hidrogenoides} and \ref{entropias_hidrogenoides} for all systems of interest.

The analysis of them is organized as follows: the results for the ground state energy are discussed in subsection~\ref{sec:4.1}, while in subsection~\ref{sec:4.2} the global behavior of $S_r$ and $S_p$ are presented. In subsection~\ref{sec:4.3} the results of $S_t$ are summarized with the discussion of the entropic uncertainty relation, suggestion of conjectures and use of $S_t$ as a measure of basis function set quality in confined environments. Finally we have examined the limit for the strong confinement regime in subsection~\ref{sec:4.4}.

\begin{table}
\scriptsize
\caption{Energy $E$ and entropy sum $S_t$ values as a function of the confinement radius $r_c$ for the confined H$_c$, He$_c^{+}$ e Li$_c^{2+}$ in the ground states. All values are in atomic units (a.u.).}
\label{energias_soma_entropica_hidrogenoides}
\begin{tabular}{ccccccccccc}
\\ \hline \hline
   $r_c$ & & & &	Ref.\cite{entropiaatomosconfinados} &   &$\phi_{11}^c(r)$  &	$\phi_{12}^c(r)$  &	$\phi_{13}^c(r)$ & $\phi_{2}^c(r)$ & $\phi_3^c(r)$
\\ \hline 
&& & & \multicolumn{7}{c}{H$_c$} 
\\
0.5 & & $E$	& &	14.748	& &  14.8970	&14.8152	&14.8774		&14.8316& \textbf{\textit{14.7644}}
	\\
& & $S_t$		& & 6.497 	& & 6.5794	&6.5766	&6.5803		&6.5779		&\textbf{\textit{6.5759}}
\\  
1.0	& &$E$		& &2.374 & &	2.3906		&2.3784&	2.3878	&	2.3806		&\textbf{\textit{2.3741}}
\\
	& &$S_t$		& &6.520	& &	6.5449 &	6.5413		&6.5458
&	6.5426	&\textbf{\textit{6.5395}}
\\  
1.5	& &$E$& &		0.437	&  &0.4388	&\textbf{\textit{0.4371}}&	0.4384	&	0.4371	& 0.4389
\\
	& &$S_t$	& &	6.506	& &	6.5157	&\textbf{\textit{6.5110}}	&6.5161
&		6.5125	&	6.5077
\\  
2.0	& &$E$& &	-0.125 & 	&\textbf{\textit{-0.1250}}	&-0.1240	&-0.1249	&-0.1245	&-0.1211
	\\
	& &St& &		6.495 & &	 \textbf{\textit{6.4960}}	&6.4900	&6.4958& 	6.4918	&	6.4853
 \\  
2.5	& &$E$& &		-0.335 & &	\textbf{\textit{-0.3343}}	&		-0.3325	&		-0.3342	&	-0.3332	&	-0.3296
 \\
 & &	$S_t$& &		6.494 & &	\textbf{\textit{6.4859}}	&		6.4794&			6.4857	&	6.4813	&	6.4740
 \\  
3.0	& &$E$& &		-0.424 & & \textbf{\textit{-0.4225}}	&-0.4206	&-0.4224		&-0.4213		&-0.4180
 	\\
 & &	$S_t$& &		6.499 & &	\textbf{\textit{6.4838}}	&6.4771	&6.4837&		6.4792	&6.4715
 \\  
3.5	& &$E$& &		-0.464 		& & \textbf{\textit{-0.4624}}	&-0.4606	&-0.4623		&-0.4612		&-0.4584
 \\
 & &	$S_t$& &		6.509 & &	\textbf{\textit{6.4879}}	&6.4813	&6.4880	&	6.4833	&	6.4758
 \\  
4.0	& &$E$	& &		-0.483 & &	-0.4811	&-0.4796	&\textbf{\textit{-0.4811}}		&-0.4801	&	-0.4779
 	\\
 & &	$S_t$	& &	6.521 & &	6.4960	&6.4895	&\textbf{\textit{6.4964}}&	6.4916	&6.4843
 \\  
4.5& &	$E$& &		-0.492 & & -0.4902	&-0.4889&	\textbf{\textit{-0.4902}}		&-0.4894		&-0.4876
\\
& &$S_t$& &		6.533& &	 6.5058&	6.4995&	\textbf{\textit{6.5065}}	&	6.5016	&	6.4944
\\  
5.0	& &$E$& &		-0.496 & &-0.4947	&-0.4937	&\textbf{\textit{-0.4948}}&		-0.4941&		-0.4927
\\
& &$S_t$& &		6.544 & &	6.5156	&6.5094	&\textbf{\textit{6.5165}}	&	6.5116	&	6.5045
 \\  
5.5	& &$E$& &		-0.498 & &	-0.4970	&-0.4962	&\textbf{\textit{-0.4971}}		&-0.4965	&-0.4
955
\\
	& &$S_t$	& &		6.552	& &		6.5244	&6.5184&	\textbf{\textit{6.5255}	}	&	6.5206		&	6.5138
\\  
6.0	& &$E$& &		-0.499 & &	-0.4982	&-0.4976	&\textbf{\textit{-0.4983}}	&	-0.4979	&-0.4971
\\
& &$S_t$& &		6.557& &	6.5312&	6.5260	&\textbf{\textit{6.5331}}	&	6.5282&	6.5216
\\ \hline
&& & & \multicolumn{7}{c}{He$_c^{+}$}
\\  
0.5& &	$E$	& &	9.496	& &	9.5623	&9.5135&	9.5513	&	9.5222	&	\textbf{\textit{9.4965}}
\\
& &$S_t$& &		6.525		& & 6.5449	&6.5414	&6.5458&	6.5427	&	\textbf{\textit{6.5394}}
\\  
1.0	& &$E$& &		-0.500	& &	\textbf{\textit{-0.5000}}	&-0.4959	&-0.4996	&	-0.4979	&	-0.4842
\\
& &$S_t$	& &	6.498& &	\textbf{\textit{6.4959}}	&6.4900	&6.4959	&	6.4918		&6.4852

\\  
1.5& &	$E$& &		-1.696& &	\textbf{\textit{-1.6902}}	&-1.6823	&-1.6897		&-1.6850		&-1.6721
\\
& &	$S_t$& &			6.496	& &	\textbf{\textit{6.4838}}	&6.4771&	6.4837	&	6.4791	&		6.4715
\\  
2.0	& &$E$& &		-1.933& &	-1.9244	&-1.9182	&\textbf{\textit{-1.9245}}	&	-1.9204		&-1.9115
\\
& &	$S_t$	& &	6.517		& & 6.4960	&6.4895	&\textbf{\textit{6.4964}}	&	6.4916		&6.4842
\\  
2.5	& &$E$& &		-1.985& &	-1.9787	&-1.9747	&\textbf{\textit{-1.9790}}	&	-1.9762	&	-1.9708
\\
& &$S_t$& &		6.541& &		6.5157	&6.5095	&\textbf{\textit{6.5165}}	&	6.5116		&6.5046
\\  
3.0	& &$E$& &		-1.997& &	-1.9929	&-1.9905	&\textbf{\textit{-1.9933}}	&-1.9915	&	-1.9883
\\
& &	$S_t$& &		6.555& &	6.5318	&6.5260	&\textbf{\textit{6.5332}}&		6.5282	&	6.5216
\\  
3.5	& &$E$& &		-1.999& &	-1.9972	&-1.9957	&\textbf{\textit{-1.9975}}		&-1.9964	&-1.9945
\\
& &St& &		6.563& &	6.5425	&6.5373	&\textbf{\textit{6.5442}}	 &	6.5394	&	6.5334
\\  
4.0& &	$E$& &		-1.999& & -1.9987	&-1.9978	&\textbf{\textit{-1.9989}}	&-1.9982		&-1.9970
\\
& &$S_t$& &		6.565& &		6.5494	&6.5447	&\textbf{\textit{6.5512}}	&	6.5467	&	6.5415
\\  
4.5& &	$E$		& &-2.000	& &	-1.9993	&-1.9987	&\textbf{\textit{-1.9995}}		&-1.9990		&-1.9983
\\
& &$S_t$		& &6.566& &		6.5538	&6.5498	&\textbf{\textit{6.5557}}&		6.5516		&6.5470
\\  
5.0	& &$E$	& &	-2.000& &	-1.9996	&-1.9992	&\textbf{\textit{-1.9997}}	&	-1.9994	&	-1.9989
\\
& &$S_t$& &		6.566& &	6.5569	&6.5532&	\textbf{\textit{6.5586}}	&	6.5548	&	6.5509
\\ \hline
&& & & \multicolumn{7}{c}{Li$_c^{2+}$}
\\   
0.5 & &	$E$ & &		3.933 & &	3.9495	&\textbf{\textit{3.9338}}&	3.9456	&	3.9343	 &	3.9498
	\\
 & &	$S_t$ & &		6.505 & &	6.5158	&\textbf{\textit{6.5110}}	&6.5161	 &	6.5125	 &	6.5077
\\  
1.0 & &$E$		& &-3.816& & 	\textbf{\textit{-3.8029}}	&-3.7853	&-3.8018	&	-3.7913		&-3.7622
\\
	& &$S_t$& &		6.493& &		\textbf{\textit{6.4838}}	&6.4771	&6.4837	&	6.4791	&6.4715
\\  
 1.5	& &$E$& &		-4.430& & 	-4.4114	&-4.4002	&\textbf{\textit{-4.4120}}	&	-4.4043	&	-4.3885
 \\
 	& &$S_t$& &		6.527		& &	6.5058	&6.4995	&\textbf{\textit{6.5065}}	&	6.5016		&6.4944
\\  
2.0& &	$E$& &		-4.493& & 	-4.4839	&-4.4786	&\textbf{\textit{-4.4849}}	&	-4.4808		&-4.4736
\\
& & $S_t$& &		6.554	& &	6.5318	&6.5260	&\textbf{\textit{6.5331}}	&	6.5281		&6.5216
\\  
2.5	& &$E$& &		-4.500& & 	-4.4957	&-4.4932	&\textbf{\textit{-4.4964}	}	&-4.4943		&-4.4910
\\
& &$S_t$& &		6.563& &		6.5464	&6.5414	&\textbf{\textit{6.5481}}&	6.5435	&	6.5378
\\  
3.0& &	$E$& &		-4.500	& &	-4.4984	&-4.4972	&\textbf{\textit{-4.4989}}	&	-4.4978		&-4.4961
\\
& &$S_t$	& &	6.564	& &	6.5539	&6.5498	&\textbf{\textit{6.5557}}	&	6.5516	&	6.5470
\\  
3.5	& &$E$		& &-4.500		& & -4.4993	&-4.4986	&\textbf{\textit{-4.4996}}	&	-4.4990	&	-4.4980
\\
& &$S_t$		& &6.566	& &	6.5580	&6.5546	&\textbf{\textit{6.5597}}	&	6.5561		&6.5524
\\ \hline \hline
\end{tabular}
\end{table}

\begin{table}
\scriptsize  
\caption{The Shannon entropies $S_r$ e $S_p$ as a function of the confinement radius $ r_c $ for the confined H$_c$, He$_c^{+}$ e Li$_c^{2+}$ in the ground state. All values are in atomic units (a.u.).}
\label{entropias_hidrogenoides}
\begin{tabular}{ccccccccccc}
\\ \hline \hline
   $r_c$ & & & &	Ref.\cite{entropiaatomosconfinados} &   &$\phi_{11}^c(r)$  &	$\phi_{12}^c(r)$  &	$\phi_{13}^c(r)$ & $\phi_{2}^c(r)$ & $\phi_3^c(r)$
\\ \hline 
&& & & \multicolumn{7}{c}{H$_c$} 
\\
0.5& & $S_r$ & &	-1.470 & & -1.4537	&-1.4598	&-1.4531	& -1.4577		&-1.4652 \\
   & & $S_p$ & &	7.967  & &  8.0331	&8.0364&	8.0334	&	8.0356		& 8.0411 \\  
1.0& & $S_r$ & &  0.529  & & 0.5411	&0.5344&	0.5417	&	0.5366		& 0.5283 \\
   & & $S_p$ & &	5.991 &  & 6.0038 	&6.0069 &6.0041	&	6.0060		& 6.0112 \\ 
1.5& & $S_r$ & &  1.649 &  & 1.6559	&1.6481	&1.6564	&	1.6506		& 1.6408 \\
   & & $S_p$ & & 4.857 & & 4.8598	&4.8629	& 4.8597 &	4.8619		& 4.8669 \\  
2.0& & $S_r$ & & 	2.397& & 2.3967	&2.3872	&2.3970		& 2.3902		& 2.3783 \\
   & & $S_p$ & & 4.099 & & 4.0993	&4.1028	&4.0988	&	4.1016		& 4.1070 \\  
2.5& & $S_r$ & & 2.929& & 2.9198	&2.9081&	2.9200		& 2.9117		& 2.8969 \\
   & & $S_p$ & & 3.565& & 3.5661	&3.5713	&3.5657	&	3.5696		& 3.5771 \\  
3.0& & $S_r$ & & 3.316&	& 3.2944	&3.2800	&3.2947	&	3.2845	& 3.2661 \\
   & & $S_p$ & & 3.183& & 3.1894	&3.1971	&3.1891	& 3.1947	& 3.2054 \\  
3.5& & $S_r$ & & 3.595&	& 3.5594	&3.5421	&3.5600	&	3.5476	&	3.5259 \\
   & & $S_p$ & & 2.914& & 2.9285	&2.9392	&2.9280	&	2.9357	&2.9499 \\  
4.0& & $S_r$ & & 3.791&	& 3.7422	&3.7225	&3.7435	&	3.7290	& 3.7047 \\
   & & $S_p$ & & 2.730& & 2.7538	&2.7670	&2.7529	&	2.7626	& 2.7796 \\  
4.5& & $S_r$ & & 3.925&	& 3.8650	&3.8440	&3.8673	&	3.8512	&3.8255 \\
   & & $S_p$ & & 2.609& & 2.6408	&2.6555	&2.6392	&	2.6505	& 2.6689 \\  
5.0& & $S_r$ & & 4.011&	& 3.9463	&3.9249	&3.9496	&	3.9326	& 3.9069 \\
   & & $S_p$ & & 2.533& & 2.5693	&2.5845	&2.5669	& 2.5791	& 2.5976 \\  
5.5& & $S_r$ & & 4.064&	& 4.0000  &3.9792	&4.0043	& 3.9869	&	3.9623 \\
   & & $S_p$ & & 2.487& & 2.5244	&2.5392	&2.5212	&	2.5337	& 2.5515 \\  
6.0& & $S_r$ & & 4.095& & 4.0354	&4.0163	&4.0410	&	4.0239	& 4.0008 \\
   & & $S_p$ & & 2.462& & 2.4958 	&2.5097	&2.4921	&	2.5043	& 2.5208 
\\ \hline
&& & & \multicolumn{7}{c}{He$_c^{+}$}
\\  
0.5 & & $S_r$ & & -1.550& &-1.5383	&-1.5450	&-1.5377	& -1.5428		& -1.5512 \\
    & & $S_p$ & & 8.075& &8.0832	&8.0864 &8.0835	&	8.0855	& 8.0906 \\  
1.0	& & $S_r$ & & 0.317& & 0.3172	&0.3078	&0.3176		& 0.3108		& 0.2988 \\
    & & $S_p$ & & 6.180& & 6.1787	&6.1822	&6.1783	&	6.1810	&6.1864 \\  
1.5	& & $S_r$ & & 1.237& & 1.2150	&1.2005	&1.2152	& 1.2050	& 1.1867 \\
    & & $S_p$ & & 5.259& & 5.2688	&5.2766	&5.2685	&	5.2741	& 5.2848 \\  
2.0	& & $S_r$ & & 1.714& & 1.6627	&1.6431	&1.6641	&	1.6495	&1.6252 \\
    & & $S_p$ & & 4.804& & 4.8333	&4.8464	&4.8323	&	4.8421 & 4.8590 \\  
2.5	& & $S_r$ & & 1.937& & 1.8669	&1.8455	&1.8702	&1.8531		&1.8275 \\ 
    & & $S_p$ & & 4.604& & 4.6488	&4.6640	&4.6463	&	4.6585	&4.6771 \\  
3.0	& & $S_r$ & & 2.023& & 1.9566	&1.9368	&1.9616	&1.9444	&1.9213 \\
    & & $S_p$ & & 4.532& &4.5752	&4.5892	&4.5716	&	4.5838	&4.6003 \\  
3.5	& & $S_r$ & & 2.055& &1.9985	&1.9815	&2.0043	&	1.9885	&1.9689 \\
    & & $S_p$ & & 4.508& &	4.5440	&4.5558	&4.5399	&	4.5509	&4.5645 \\  
4.0	& & $S_r$ & & 2.063& &2.0203	&2.0060	&2.0263	&	2.0122	&1.9959 \\
    & & $S_p$ & & 4.502& & 4.5291	&4.5387	&4.5249		&4.5345		&4.5456 \\  
4.5	& & $S_r$ & & 2.065& &2.0329	&2.0209	&2.0387	&	2.0263	&2.0127 \\
    & & $S_p$ & & 4.501& &4.5209	&4.5289	&4.5170	& 4.5253	&4.5343 \\  
5.0	& & $S_r$ & & 2.065& &2.0409	&2.0306	&2.0463	&	2.0353	&2.0239 \\
    & & $S_p$ & & 4.501& &4.5160	&4.5226	&4.5123		& 4.5195	&4.5270 
	\\ \hline
&& & & \multicolumn{7}{c}{Li$_c^{2+}$}
\\ 
0.5& & $S_r$ & & -1.650& &	-1.6399	&-1.6477	&-1.6394	&	-1.6452	&-1.6550 \\
    & & $S_p$ & & 8.152& &	8.1557	&8.1587	&8.1555	&	8.1577		&8.1627 \\  
1.0& & $S_r$ & & 0.021& & -0.0014	&-0.0159	&-0.0012		&-0.0114	& -0.0297 \\
    & & $S_p$ & & 6.472& & 6.4852	&6.4930	&6.4849	&6.4905	& 6.5012 \\  
1.5& & $S_r$ & & 0.633& & 0.5692	&0.5481	&0.5715	&	0.5553	&0.5297 \\
    & & $S_p$ & & 5.894& & 5.9366	&5.9514	&5.9350	&	5.9463	&5.9647 \\  
2.0& & $S_r$ & & 0.808& & 0.7402 &0.7204	&0.7452	&	0.7280	&0.7049 \\
    & & $S_p$ & & 5.745& & 5.7916	&5.8056&	5.7879	&	5.8001	&5.8167 \\  
2.5& & $S_r$ & & 0.843& & 0.7946	&0.7790	&0.8006	&	0.7856		&0.7677 \\
    & & $S_p$ & & 5.719& & 5.7518	&5.7624	&5.7475	&	5.7579	&5.7701 \\  
3.0& & $S_r$ & & 0.848& & 0.8165	&0.8045	&0.8223	&	0.8099	&0.7963 \\
    & & $S_p$ & & 5.716& & 5.7373	&5.7453	&5.7334	&	5.7417	&5.7507 \\  
3.5& & $S_r$ & & 0.849& & 0.8274	&0.8179	&0.8326	&0.8223	&0.8117 \\
    & & $S_p$ & & 5.717& &5.7306	&5.7367	&5.7271	&	5.7338		&5.7407 \\ 
		\hline  \hline
\end{tabular}
\end{table}

\subsection{Energy}\label{sec:4.1}

\begin{figure}[h]
\centering 
\includegraphics[width=9cm, height=7cm]{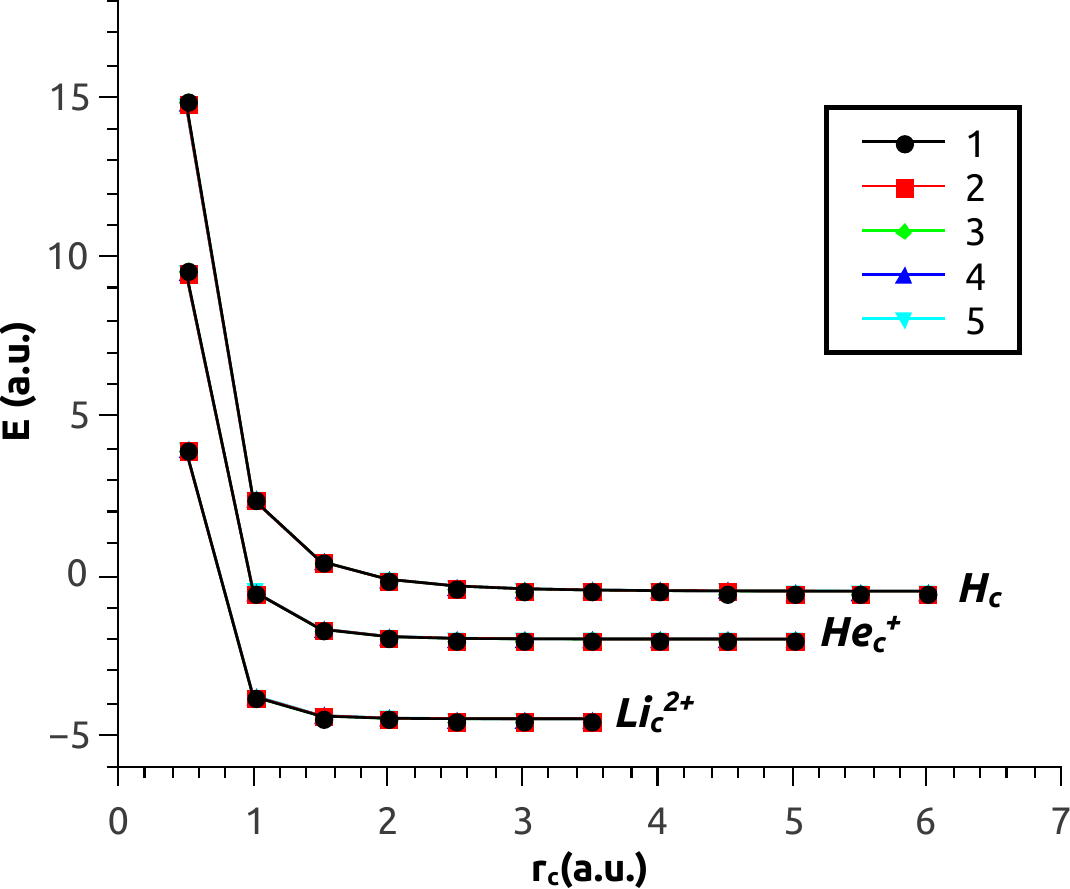}
\caption{Ground state energy $E$ as a function of the confinement radius $r_c$ for the confined H$_c$, He$_c^{+}$ and Li$_c^{2+}$. All values are in atomic units (a.u.). Subtitles: 1=$\phi_{11}^c(r)$, 2=$\phi_{12}^c(r)$, 3=$\phi_{13}^c(r)$, 4=$\phi_{2}^c(r)$ and 5=$\phi_{3}^c(r)$.} 
\label{energias_hidrogenoides} 
\end{figure}

The ground state energy values of confined H$_c$, He$_c^{+}$ e Li$_c^{2+}$ for each type of trial wave functions and different values of $r_c$ are presented in Table \ref{energias_soma_entropica_hidrogenoides}, and their general behaviors are displayed in Figure \ref{energias_hidrogenoides}. In general, the energy values of these confined systems tend to the ones of the free systems when $r_c$ goes to infinity and increases when $r_c$ decreases, reaching positive values for small $r_c$. In addition, one can clearly see that the one electron atomic system with the highest atomic number has lower energies for the same confinement radius.

Analyzing the quality of the trial wave functions considered here by using the energy variational criterion, we can note that not necessarily the same trial wave function provides the best values for all $r_c$ studied. In Table~\ref{energias_soma_entropica_hidrogenoides} we indicated in bold and italic the best variational energy value for each value of $r_c$ and the corresponding value of $S_t$. For Li$_c^{2+}$,  for example, the trial wave function $\phi_{13}^c(r)$ provides the best energy value for $r_c$ comprised between $1.5 \ a.u.$ and $3.5 \ a.u.$, while for $r_c=1.0 \ a.u. \ $ the trial wave function $\phi_{11}^c(r)$ presents the best result and, finally, for $r_c=0.5 \ a.u. \ $ the trial wave function with the best energy value is the $\phi_{12}^c(r)$.

\subsection{Shannon entropies \( S_{r} \) and \( S_{p} \)}\label{sec:4.2}

\begin{figure}[h]
\centering 
\includegraphics[width=9cm, height=7cm]{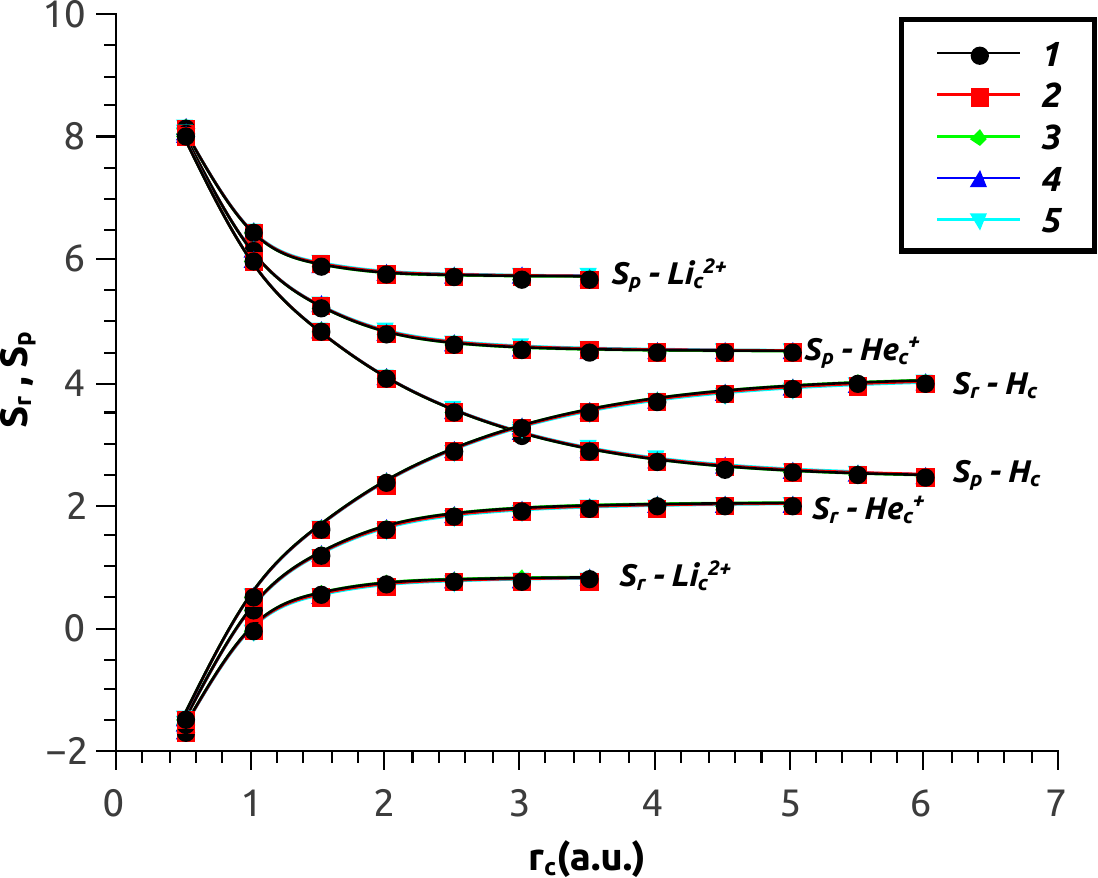}
\caption{Shannon entropies $S_r$ e $S_p$ as a function of the confinement radius $ r_c $ for the confined H$_c$, He$_c^{+}$ and Li$_c^{2+}$ in the ground states. Subtitles: 1=$\phi_{11}^c(r)$, 2=$\phi_{12}^c(r)$,	3=$\phi_{13}^c(r)$, 4=$\phi_{2}^c(r)$ and 5=$\phi_{3}^c(r)$.} 
\label{entropias_de_shannon} 
\end{figure}

The Shannon entropies in the position ($S_r$) and momentum ($S_p$) spaces of confined H$_c$, He$_c^{+}$ and Li$_c^{2+}$ in the ground state for different values of confinement radius ($ r_c $) are presented in Table~\ref{entropias_hidrogenoides}, and their general behaviors are displayed in Figure~\ref{entropias_de_shannon}. Based on such data, we have verified a decrease of $S_r$ value and an increase of $S_p$ values when $r_c$ decreases and the confinement effect is enhanced. Additionally, we can observe that $S_r$ is smaller (and $S_p$ is greater) for larger values of $Z$, that is, such quantities have a dependence on atomic nucleus number, corroborating with the analytical work presented in Ref.~\cite{shannon_correlacao}. These results evidence the interpretation that $S_r$ indicates a measure of the uncertainty in the spatial location of the particle, and the global behavior of the quantities $S_r$ and $S_p$ are justifiable considering Eq.~(\ref{soma_entropica}). An interesting result for H$_c$ case is the intersection of $S_r$ and $S_p$ curves to $r_c \approx 3.0$ a.u.. At this point $S_r$ and $S_p$ assume the value near to 3.25.      
 
Moreover, the numerical values of $S_r$ (and $S_p$) for these three hydrogenic-like atoms approximate when the confinement becomes more intense. In the region where the confinement effect is rigorous the quantity $S_r $ may assume negative values, as shown in Table~\ref{entropias_hidrogenoides} and in Refs.~\cite{entropiaatomosconfinados,wallas_fred_educacional, benchmark_hidrogenio}. This result has a simple explanation in the quantum context~\cite{aquino}: when $r_c$ is reasonably small, the probability density becomes large and $a_0^n\rho(\vec{r}) > 1$. In this situation, $-\rho(\vec{r}) \ln a_0^n\rho(\vec{r}) < 0$ and, then, $S_r$ may be negative. Remember that the wave-function normalization is in relation to the integral $\int \rho(\vec{r}) d\vec{r}$. Note that in the original Shannon's work the entropy values for continuous distributions may be negative (see pg. 631 in Ref.~\cite{shannonoriginal1}).   

\subsection{The entropy sum}\label{sec:4.3}

\begin{figure}[h] 
\centering
\includegraphics[width=9cm, height=7cm]{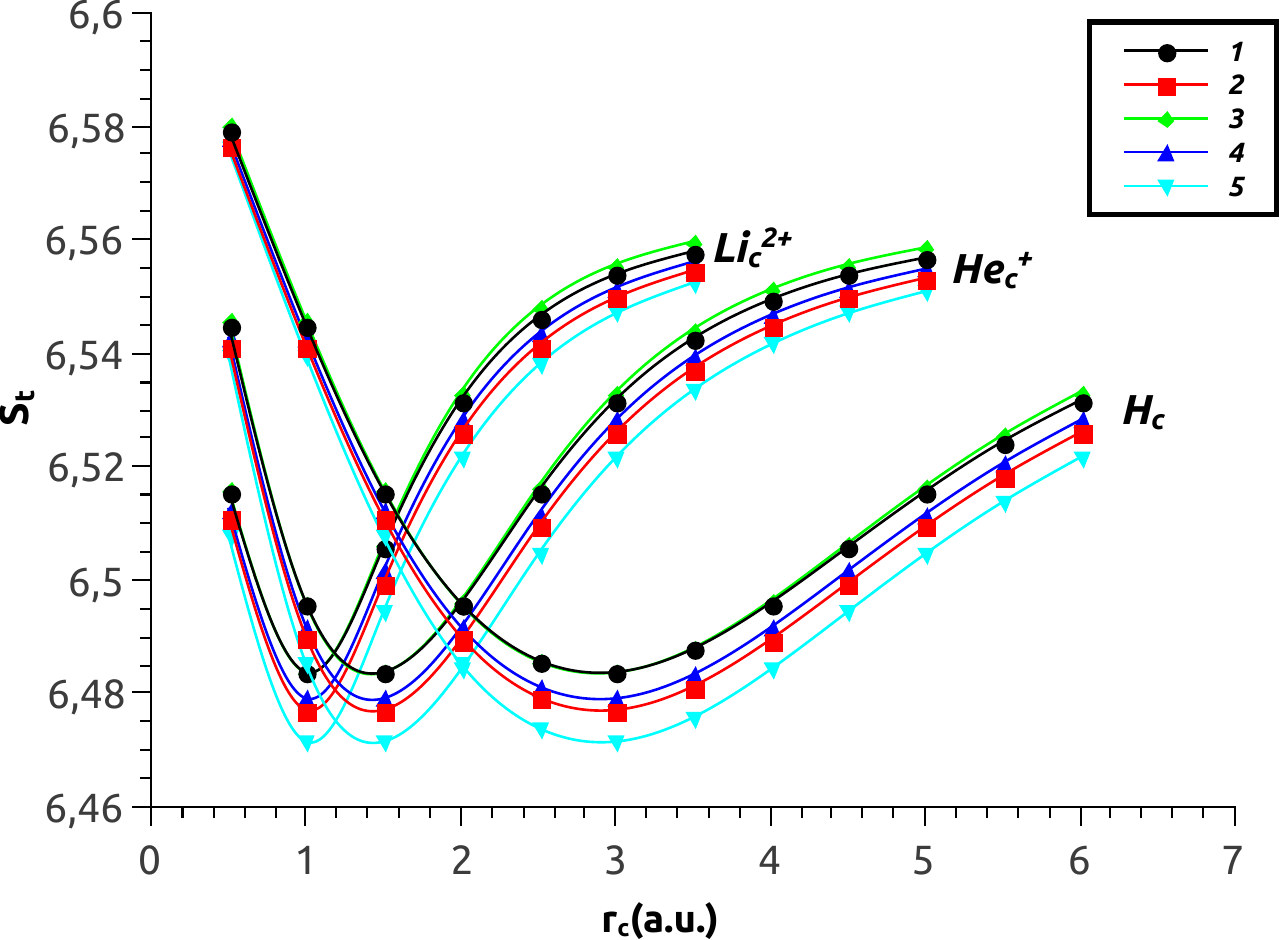}
\caption{Entropy sum $S_t$ as a function of the confinement radius $r_c$ for the confined 1H$_c$, He$_c^{+}$ and Li$_c^{2+}$ in the ground states. Subtitles: 1=$\phi_{11}^c(r)$, 2=$\phi_{12}^c(r)$,	3=$\phi_{13}^c(r)$, 4=$\phi_{2}^c(r)$ and 5=$\phi_{3}^c(r)$.}
\label{somaentropica} 
\end{figure}

The entropy sum ($S_t$), from the $S_r$ and $S_p$ values presented in Table~\ref{entropias_hidrogenoides}, of confined H$_c$, He$_c^{+}$ and Li$_c^{2+}$ in the ground state for different values of confinement radius ($ r_c $) and trial wave functions are presented in Table~\ref{energias_soma_entropica_hidrogenoides}, and their general behavior are displayed in Figure~\ref{somaentropica}. An immediate observation is that the entropic uncertainty relation ($S_t \ge 6.4342$) is valid for all studied confined systems. Still, based on this Table, we conjecture that the values of $S_t$ for confined H-like atoms considered here retain a  dependence on the atomic number $Z$ and are invariant through of the relation $\frac{r_c}{Z}$. For the three free hydrogenic-like systems the value of entropy sum is invariant ($S_t=6.5666$), corroborating with the analytical study presented in Ref.~\cite{shannon_correlacao}. 

Moreover, we can identified in the Figure~\ref{somaentropica} that $S_t$ assumes a minimum value at $r_{c}\approx 1.0 \ a.u.$,  $r_c\approx 1.5 \ a.u.$ and $r_c\approx 3.0 \ a.u.$ for the Li$_c^{2+}$, He$_c^{+}$ and H$_c$, respectively. This is not observed in Ref.~\cite{entropiaatomosconfinados} to H$_c$ case, where the authors found that the minimal value of $S_t$ is to $r_c \approx 2.5 \ a.u.$. However, we have believed in some kind of numerical problem in those results. Note from present results that the $r_c$ value when the entropy sum is minimal decreases with increasing of the atomic number, maintaining the following relationship: $r_c=\frac{3}{Z}$. For this confinement radius, the Coulomb and the confining potentials are compensated by creating a state of lower entropic uncertainty.

Additionally, we study the Gadre-Sears-Chakravorty-Bendale (GSCB) conjecture, originally analyzed for free neutral atoms, in confined environments and also to ionized atoms, from results presented on Table~\ref{energias_soma_entropica_hidrogenoides}. In particular, the GSCB conjecture establishes that better basis functions (smaller values for energy) present higher values for the entropic sum. In relation to H$_c$, the wave function that provides the best value of $E$ also provides the largest value of $S_t$ to $r_c$ values between $2.0 \ a.u.$ and $6.0 \ a.u.$. There is an exception at $r_c=3.5 \ a.u.$ where there is a inversion between $\phi_{11}^c(r)$ and $\phi_{13}^c(r)$ wave functions  but the results are identical up to four figures. Similar results are observed for He$_c^{+}$ and Li$_c^{2+}$ to $r_c$ values between $1.0 \ a.u.$ and $5.0 \ a.u.$ and $1.0 \ a.u.$ and $3.5 \ a.u.$, respectively. However, the GSCB conjecture fails when the ground state energy of the confined H-like system is positive, and, for various $r_c$ values, the best wave function (according to energy criteria) comes from the lowest value of $S_t$. 

\subsection{Strong confinement regime}\label{sec:4.4}

In previous study of confined harmonic oscillator~\cite{wallas_fred_educacional}, we proposed that the entropy sum can be employed to determine a strong or rigorous confinement regime. This regime is defined when the influence of the confining potential becomes greater than the free-system potential to specific configurations (or $r_c$ values). 

To verify this proposal in the present case studies, we have calculated for $r_c$ values between $0.01 \ a.u.$ and $0.4 \ a.u.$ the $S_t$ of confined H$_c$, He$_c^{+}$ and Li$_c^{2+}$ in the ground state, comparing with the results for an electron confined in an impenetrable spherical cage ($e^-_c$ system). In particular, we have utilized here only the $\phi_{3}^c(r)$ trial wave function to describe the confined H-like systems because it have showed the best behavior for small $r_c$ values, while the exact solution for the ground state $e^-_c$ system can be found in several textbooks (see, {\it e. g.}, Ref.~\cite{griffiths}). 

\begin{figure}[h] 

\centering 
\includegraphics[width=9cm, height=7cm]{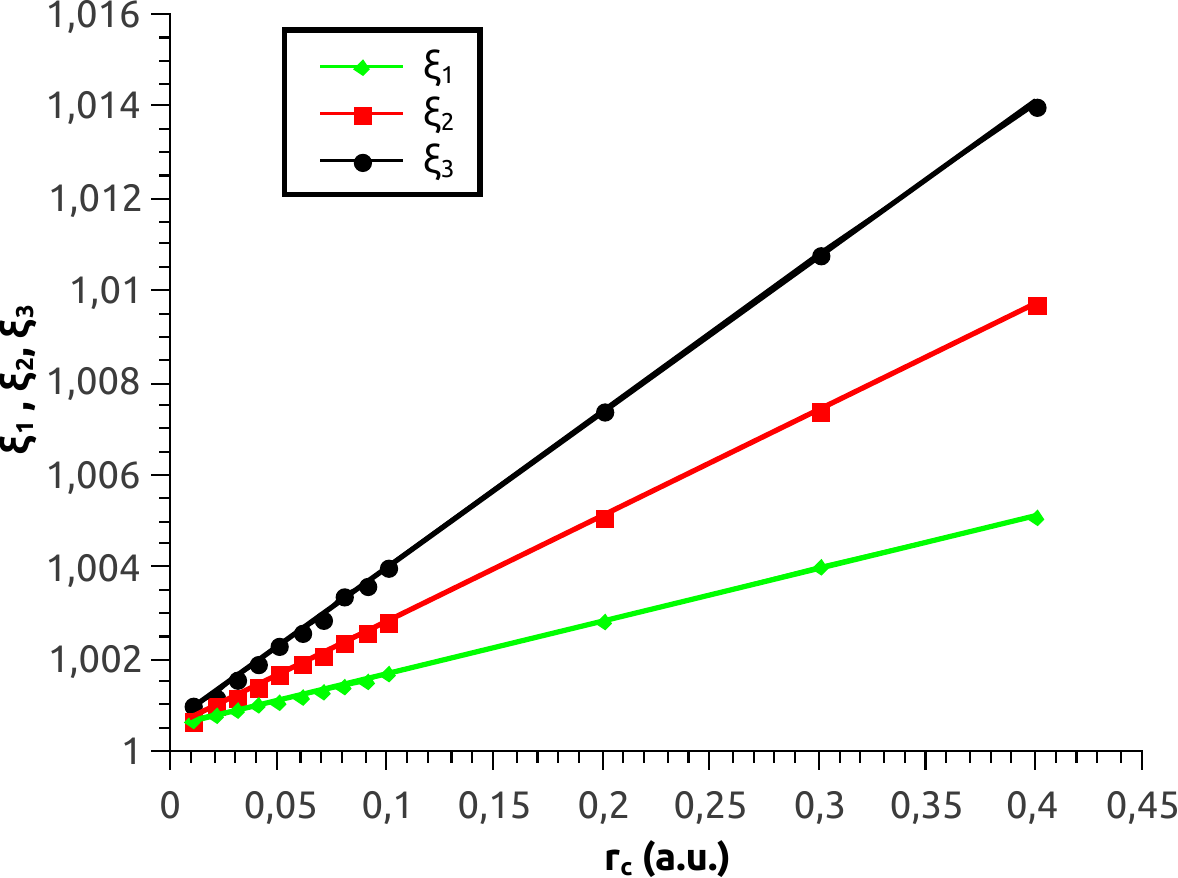}
\caption{Linear regressions of the ratios $\xi_1$, $\xi_2$ and $\xi_3$ as a function of the confinement radius $r_c$ .}
\label{comparacao_confinamento_rigoroso}
\end{figure}

The ratios between $S_t$ values of $e^-_c$ ($S_t(e^-_c)$) and of confined H-like systems ($S_t({\rm H}_c)$, $S_t({\rm He}^+_c)$ and $S_t({\rm Li}^{2+}_c)$) are displayed in Figure~\ref{comparacao_confinamento_rigoroso}, where  $\xi_1=\frac{S_t(e^-_c)}{S_t({\rm H}_c)})$, $\xi_2 =\frac{S_t(e^-_c)}{S_t({\rm He}^+_c)})$ and $\xi_3 =\frac{S_t(e^-_c)}{S_t({\rm Li}^{2+}_c)}$. In such a Figure, we have observed that these differents $S_t$ tends to the one of $e^-_c$, that is $S_t(e^-_c)=6.6172$. For example, at $r_c=0.01 \ a.u.$, $S_t({\rm H}_c)=6.6127$, $S_t({\rm He}^+_c)=6.6123$ and $S_t({\rm Li}^{2+}_c)=6.6106$. Thus, we can consider that to $r_c$ values smaller than $0.1 \ a.u.$ the confined systems are certainly in a strong or rigorous confinement regime, and the effect of the Coulomb potential is supplanted by the impenetrable spherical cage potential. The present results and the analyzes made in Ref.\cite{wallas_fred_educacional} respond to the recent questions presented in Ref.\cite{benchmark_hidrogenio} with respect to the behavior of entropy sum when the confinement radius tends to zero.

\section{Conclusion}\label{sec:5}

In the present work a variational study on one-electron atomic systems (hydrogen, helium cation and lithium dication) confined within impenetrable walls was performed by using the Shannon entropies and entropy sum. Novel expressions to the entropy sum $S_t$ and Shannon entropies $S_r$ and $S_p$ were proposed to ensure their dimensionless characteristic. Subsequently, the energy, $S_r$, $S_p$ and $S_t$ of H$_c$, He$_c^{+}$ and Li$_c^{2+}$ in their ground state were calculated for different confinement radii by using different trial wave functions. In particular, such trial functions provide reasonable results with the computed energy values close to the results already presented in the literature. 

The global behavior of these quantities were analyzed and two conjectures were tested. The first conjecture establishes that $S_t$ values for confined H-like systems are invariant through of the relation $\frac{r_c}{Z}$, while the second one determines that to $r_c=r_c^{mim}\equiv \frac{3.0}{Z} \ a.u.$ the entropy sums $S_t({\rm H}_c)$, $S_t({\rm He}^+_c)$ and $S_t({\rm Li}^{2+}_c)$ assume their minimum values. Specifically, when the confinement radius is $r_c^{mim}$, the Coulomb and the confining potentials are compensated by creating a state of lower entropic uncertainty, while for $r_c \gg r_c^{mim}$  the Coulomb potential becomes the most relevant. For $r_c \ll r_c^{mim}$ the confined H-like systems are in  a strong or rigorous confinement regime when the effect of the impenetrable spherical cage potential becomes the most relevant. This indicates that the values of $S_t$ for the confined H$_c$, He$_c^{+}$ and Li$_c^{2+}$ tends to the values obtained for an electron confined in an impenetrable spherical cage ($e^-_c$ system), both in the ground state.

In addition, we verified that the entropic uncertainty relation is valid for all studied confined systems. However, the GSCB conjecture that defines $S_t$ as a measure of trial wave functions quality has only been verified in confined environments for confinement radius whose ground state energies are negative. When this energy becomes positive, the GSCB conjecture fails and, for various $r_c$, the best trial wave function comes from the lowest value of entropy sum. 

As a further application we plan to study the confined He-like systems by using the Shannon entropies and the entropy sum within a variational framework.

\section*{Acknowledgments}

This work has been supported by the Brazilian agencies CAPES (Coordena\c{c}\~ao de Aperfei\c{c}oamento de Pessoal de N\'ivel Superior) and CNPq (Conselho Nacional de Desenvolvimento Cient\'ifico e Tecnol\'ogico) through grants to the authors. The authors thank the referees for careful reading of the manuscript and for helpful comments and suggestions.

\section*{References}

\bibliography{ms}

\begin{thebibliography}{50}
\expandafter\ifx\csname natexlab\endcsname\relax\def\natexlab#1{#1}\fi
\providecommand{\url}[1]{\texttt{#1}}
\providecommand{\href}[2]{#2}
\providecommand{\path}[1]{#1}
\providecommand{\DOIprefix}{doi:}
\providecommand{\ArXivprefix}{arXiv:}
\providecommand{\URLprefix}{URL: }
\providecommand{\Pubmedprefix}{pmid:}
\providecommand{\doi}[1]{\href{http://dx.doi.org/#1}{\path{#1}}}
\providecommand{\Pubmed}[1]{\href{pmid:#1}{\path{#1}}}
\providecommand{\bibinfo}[2]{#2}
\ifx\xfnm\relax \def\xfnm[#1]{\unskip,\space#1}\fi
\bibitem[{Shannon(1948)}]{shannonoriginal1}
\bibinfo{author}{C.~Shannon}, \bibinfo{journal}{Bell Syst. Tech. J.}
  \bibinfo{volume}{27} (\bibinfo{year}{1948}) \bibinfo{pages}{379--423 and
  623--656}. \DOIprefix\doi{10.1002/j.1538-7305.1948.tb01338.x}.
\bibitem[{Shannon and
  Weaver(1949)}]{livro_teoria_matemcatica_da_comunicacao_original}
\bibinfo{author}{C.~E. Shannon}, \bibinfo{author}{W.~Weaver},
  \bibinfo{title}{The Mathematical Theory of Communication},
  \bibinfo{publisher}{Illini Books}, \bibinfo{address}{Illinois},
  \bibinfo{year}{1949}.
\bibitem[{Nielsen and Chuang(2000)}]{nielsen}
\bibinfo{author}{M.~A. Nielsen}, \bibinfo{author}{I.~L. Chuang},
  \bibinfo{title}{Quantum Computation and Quantum Information},
  \bibinfo{publisher}{Cambridge University Press},
  \bibinfo{address}{Cambridge}, \bibinfo{year}{2000}.
\bibitem[{Sen(2011)}]{livro_sen_shannon}
\bibinfo{editor}{K.~D. Sen} (Ed.), \bibinfo{title}{Statistical Complexity:
  Applications in Electronic Struture}, \bibinfo{publisher}{Springer},
  \bibinfo{address}{Dordrecht}, \bibinfo{year}{2011}.
  \DOIprefix\doi{10.1007/978-90-481-3890-6}.
\bibitem[{Mukherjee et~al.(2015)Mukherjee, Roy, and Roy}]{tunelamento_shannon}
\bibinfo{author}{N.~Mukherjee}, \bibinfo{author}{A.~Roy},
  \bibinfo{author}{A.~K. Roy}, \bibinfo{journal}{Ann. Phys. (Berlin)}
  (\bibinfo{year}{2015}) \bibinfo{pages}{1--21}.
  \DOIprefix\doi{10.1002/andp.201500196}.
\bibitem[{Flores-Gallegos(2016)}]{information_thermodynamics}
\bibinfo{author}{N.~Flores-Gallegos}, \bibinfo{journal}{Chem. Phys. Lett.}
  \bibinfo{volume}{659} (\bibinfo{year}{2016}) \bibinfo{pages}{203--208}.
  \DOIprefix\doi{10.1016/j.cplett.2016.07.034}.
\bibitem[{Majern{\'i}k and Richterek(1997)}]{relacoessomaentropica2}
\bibinfo{author}{V.~Majern{\'i}k}, \bibinfo{author}{L.~Richterek},
  \bibinfo{journal}{J. Phys. A: Math. Gen.} \bibinfo{volume}{30}
  (\bibinfo{year}{1997}) \bibinfo{pages}{L49--L54}.
  \DOIprefix\doi{10.1088/0305-4470/30/4/002}.
\bibitem[{Majern{\'i}k et~al.(1999)Majern{\'i}k, Charvot, and
  Majern{\'i}kov{\'a}}]{particula_caixa_shannon2}
\bibinfo{author}{V.~Majern{\'i}k}, \bibinfo{author}{R.~Charvot},
  \bibinfo{author}{E.~Majern{\'i}kov{\'a}}, \bibinfo{journal}{J. Phys. A: Math.
  Gen.} \bibinfo{volume}{32} (\bibinfo{year}{1999})
  \bibinfo{pages}{2207--2216}. \DOIprefix\doi{10.1088/0305-4470/32/11/013}.
\bibitem[{Majern{\'i}k and Opatrn{\'y}(1996)}]{relacao_de_incerteza_oscilador}
\bibinfo{author}{V.~Majern{\'i}k}, \bibinfo{author}{T.~Opatrn{\'y}},
  \bibinfo{journal}{J. Phys. A: Math. Gen.} \bibinfo{volume}{29}
  (\bibinfo{year}{1996}) \bibinfo{pages}{2187--2197}.
  \DOIprefix\doi{10.1088/0305-4470/29/9/029}.
\bibitem[{Choi et~al.(2011)Choi, Kim, Kim, Maamache, Menouar, and
  Nahm}]{shannonconfinadohosciladortempo}
\bibinfo{author}{J.~R. Choi}, \bibinfo{author}{M.-S. Kim},
  \bibinfo{author}{D.~Kim}, \bibinfo{author}{M.~Maamache},
  \bibinfo{author}{S.~Menouar}, \bibinfo{author}{I.~H. Nahm},
  \bibinfo{journal}{Ann. Phys. (Amsterdam, Neth.)} \bibinfo{volume}{326}
  (\bibinfo{year}{2011}) \bibinfo{pages}{1381--1393}.
  \DOIprefix\doi{10.1016/j.aop.2011.02.006}.
\bibitem[{Y{\'a}{\~n}ez et~al.(1994)Y{\'a}{\~n}ez, {Van Assche}, and
  Dehesa}]{relacao_analitica_oscilador_hidrogenio}
\bibinfo{author}{R.~J. Y{\'a}{\~n}ez}, \bibinfo{author}{W.~{Van Assche}},
  \bibinfo{author}{J.~S. Dehesa}, \bibinfo{journal}{Phys. Rev. A}
  \bibinfo{volume}{50} (\bibinfo{year}{1994}) \bibinfo{pages}{3065--3079}.
  \DOIprefix\doi{10.1103/PhysRevA.50.3065}.
\bibitem[{H{\^o} et~al.(1998)H{\^o}, Weaver, and {Smith
  Jr.}}]{excitacao_shannon}
\bibinfo{author}{M.~H{\^o}}, \bibinfo{author}{D.~F. Weaver},
  \bibinfo{author}{V.~H. {Smith Jr.}}, \bibinfo{journal}{Phys. Rev. A}
  \bibinfo{volume}{57} (\bibinfo{year}{1998}) \bibinfo{pages}{4512--4517}.
  \DOIprefix\doi{10.1103/PhysRevA.57.4512}.
\bibitem[{Chattaraj and Sarkar(2003)}]{excitacao_shannon_confinado2}
\bibinfo{author}{P.~Chattaraj}, \bibinfo{author}{U.~Sarkar},
  \bibinfo{journal}{Chem. Phys. Lett.} \bibinfo{volume}{372}
  (\bibinfo{year}{2003}) \bibinfo{pages}{805--809}.
  \DOIprefix\doi{10.1016/S0009-2614(03)00516-5}.
\bibitem[{Sa{\~n}udo and L{\'o}pez-Ruiz(2008)}]{complexidade1}
\bibinfo{author}{J.~Sa{\~n}udo}, \bibinfo{author}{R.~L{\'o}pez-Ruiz},
  \bibinfo{journal}{Phys. Lett. A} \bibinfo{volume}{372} (\bibinfo{year}{2008})
  \bibinfo{pages}{5283}. \DOIprefix\doi{10.1016/j.physleta.2008.06.012}.
\bibitem[{Esquivel et~al.(2011)Esquivel, Molina-Esp{\'i}ritu, Angulo,
  Antol{\'i}n, Flores-Gallegos, and Dehesa}]{complexidade_3}
\bibinfo{author}{R.~O. Esquivel}, \bibinfo{author}{M.~Molina-Esp{\'i}ritu},
  \bibinfo{author}{J.~C. Angulo}, \bibinfo{author}{J.~Antol{\'i}n},
  \bibinfo{author}{N.~Flores-Gallegos}, \bibinfo{author}{J.~S. Dehesa},
  \bibinfo{journal}{Mol. Phys.} \bibinfo{volume}{109} (\bibinfo{year}{2011})
  \bibinfo{pages}{2353--2365}. \DOIprefix\doi{10.1080/00268976.2011.607780}.
\bibitem[{Esquivel et~al.(2010)Esquivel, Flores-Gallegos, Iuga, Carrera,
  Angulo, and Antol{\'i}n}]{shannon_reacao_quimica}
\bibinfo{author}{R.~Esquivel}, \bibinfo{author}{N.~Flores-Gallegos},
  \bibinfo{author}{C.~Iuga}, \bibinfo{author}{E.~M. Carrera},
  \bibinfo{author}{J.~C. Angulo}, \bibinfo{author}{J.~Antol{\'i}n},
  \bibinfo{journal}{Phys. Lett. A} \bibinfo{volume}{374} (\bibinfo{year}{2010})
  \bibinfo{pages}{948--951}. \DOIprefix\doi{10.1016/j.physleta.2009.12.018}.
\bibitem[{Esquivel et~al.(2009)Esquivel, Flores-Gallegos, Iuga, Carrera,
  Angulo, and Antol{\'i}n}]{shannon_reacao_quimica_2}
\bibinfo{author}{R.~O. Esquivel}, \bibinfo{author}{N.~Flores-Gallegos},
  \bibinfo{author}{C.~Iuga}, \bibinfo{author}{E.~M. Carrera},
  \bibinfo{author}{J.~C. Angulo}, \bibinfo{author}{J.~Antol{\'i}n},
  \bibinfo{journal}{Theor. Chem. Acc.} \bibinfo{volume}{124}
  (\bibinfo{year}{2009}) \bibinfo{pages}{445--460}.
  \DOIprefix\doi{10.1007/s00214-009-0641-x}.
\bibitem[{Sen(2005)}]{entropiaatomosconfinados}
\bibinfo{author}{K.~D. Sen}, \bibinfo{journal}{J. Chem. Phys.}
  \bibinfo{volume}{123} (\bibinfo{year}{2005}) \bibinfo{pages}{074110--1 --
  074110--9}. \DOIprefix\doi{10.1063/1.2008212}.
\bibitem[{Aquino et~al.(2013)Aquino, Flores-Riveros, and Rivas-Silva}]{aquino}
\bibinfo{author}{N.~Aquino}, \bibinfo{author}{A.~Flores-Riveros},
  \bibinfo{author}{J.~F. Rivas-Silva}, \bibinfo{journal}{Phys. Lett. A}
  \bibinfo{volume}{377} (\bibinfo{year}{2013}) \bibinfo{pages}{2062--2068}.
  \DOIprefix\doi{10.1016/j.physleta.2013.05.048}.
\bibitem[{Patil et~al.(2007)Patil, Sen, Watson, and {Montgomery
  Jr.}}]{shannonconfinado2}
\bibinfo{author}{S.~H. Patil}, \bibinfo{author}{K.~D. Sen},
  \bibinfo{author}{N.~A. Watson}, \bibinfo{author}{H.~E. {Montgomery Jr.}},
  \bibinfo{journal}{J. Phys. B: At., Mol. Opt. Phys.} \bibinfo{volume}{40}
  (\bibinfo{year}{2007}) \bibinfo{pages}{2147--2162}.
  \DOIprefix\doi{10.1088/0953-4075/40/11/016}.
\bibitem[{Song et~al.(2015)Song, Sun, and
  Dong}]{entropy_for_an_infinite_circular_well}
\bibinfo{author}{X.-D. Song}, \bibinfo{author}{G.-H. Sun},
  \bibinfo{author}{S.-H. Dong}, \bibinfo{journal}{Phys. Lett. A}
  \bibinfo{volume}{379} (\bibinfo{year}{2015}) \bibinfo{pages}{1402--1408}.
  \DOIprefix\doi{10.1016/j.physleta.2015.03.020}.
\bibitem[{Jiao et~al.(2017)Jiao, Zan, Zhang, and Ho}]{benchmark_hidrogenio}
\bibinfo{author}{L.~Jiao}, \bibinfo{author}{L.~Zan},
  \bibinfo{author}{Y.~Zhang}, \bibinfo{author}{Y.~Ho}, \bibinfo{journal}{Int.
  J. Quantum Chem.} \bibinfo{volume}{117} (\bibinfo{year}{2017})
  \bibinfo{pages}{e25375--n/a}. \DOIprefix\doi{10.1002/qua.25375}.
\bibitem[{Ghosal et~al.(2016)Ghosal, Mukherjee, and
  Roy}]{oscilador_confinado_simetrico_assimetrico}
\bibinfo{author}{A.~Ghosal}, \bibinfo{author}{N.~Mukherjee},
  \bibinfo{author}{A.~K. Roy}, \bibinfo{journal}{Ann. Phys. (Berlin)}
  \bibinfo{volume}{11-12} (\bibinfo{year}{2016}) \bibinfo{pages}{796--818}.
  \DOIprefix\doi{10.1002/andp.201600121}.
\bibitem[{{Le Sech} and Banerjee(2011)}]{confinadosistemacorte}
\bibinfo{author}{C.~{Le Sech}}, \bibinfo{author}{A.~Banerjee},
  \bibinfo{journal}{J. Phys. B: At., Mol. Opt. Phys.} \bibinfo{volume}{44}
  (\bibinfo{year}{2011}) \bibinfo{pages}{105003--1 -- 105003--9}.
  \DOIprefix\doi{10.1088/0953-4075/44/10/105003}.
\bibitem[{Guimar{\~a}es and Prudente(2005)}]{marcilioartigoconfinado}
\bibinfo{author}{M.~N. Guimar{\~a}es}, \bibinfo{author}{F.~V. Prudente},
  \bibinfo{journal}{J. Phys. B: At., Mol. Opt. Phys.} \bibinfo{volume}{38}
  (\bibinfo{year}{2005}) \bibinfo{pages}{2811--2825}.
  \DOIprefix\doi{10.1088/0953-4075/38/15/017}.
\bibitem[{Jask{\'o}lski(1996)}]{jaskolski}
\bibinfo{author}{W.~Jask{\'o}lski}, \bibinfo{journal}{Phys. Rep.}
  \bibinfo{volume}{271} (\bibinfo{year}{1996}) \bibinfo{pages}{1--66}.
  \DOIprefix\doi{10.1016/0370-1573(95)00070-4}.
\bibitem[{Almeida et~al.(2005)Almeida, Guimar{\~a}es, and Prudente}]{marcos}
\bibinfo{author}{M.~M. Almeida}, \bibinfo{author}{M.~N. Guimar{\~a}es},
  \bibinfo{author}{F.~V. Prudente}, \bibinfo{journal}{Rev. Bras. Ensino Fis.}
  \bibinfo{volume}{27} (\bibinfo{year}{2005}) \bibinfo{pages}{395--405}.
  \DOIprefix\doi{10.1590/S1806-11172005000300017}.
\bibitem[{Sen(2014)}]{livro_sen_confinado}
\bibinfo{editor}{K.~D. Sen} (Ed.), \bibinfo{title}{Electronic Structure of
  Quantum Confined Atoms and Molecules}, \bibinfo{publisher}{Springer},
  \bibinfo{address}{Cham}, \bibinfo{year}{2014}.
  \DOIprefix\doi{10.1007/978-3-319-09982-8}.
\bibitem[{Cruz(2009)}]{livro_sistemas_confinados1}
\bibinfo{editor}{S.~A. Cruz} (Ed.), \bibinfo{title}{Advances in Quantum
  Chemistry, Theory of Confined Quantum Systems vol 57 e 58},
  \bibinfo{publisher}{Academic Press}, \bibinfo{address}{New York},
  \bibinfo{year}{2009}.
\bibitem[{Prudente et~al.(2005)Prudente, Costa, and Viana}]{prudente}
\bibinfo{author}{F.~V. Prudente}, \bibinfo{author}{L.~S. Costa},
  \bibinfo{author}{J.~D.~M. Viana}, \bibinfo{journal}{J. Chem. Phys.}
  \bibinfo{volume}{123} (\bibinfo{year}{2005}) \bibinfo{pages}{224701--1 --
  224701--11}. \DOIprefix\doi{10.1063/1.2131068}.
\bibitem[{Olavo et~al.(2016)Olavo, Maniero, de~Carvalho, Prudente, and
  Jalbert}]{fred_ginette}
\bibinfo{author}{L.~S.~F. Olavo}, \bibinfo{author}{A.~M. Maniero},
  \bibinfo{author}{C.~R. de~Carvalho}, \bibinfo{author}{F.~V. Prudente},
  \bibinfo{author}{G.~Jalbert}, \bibinfo{journal}{J. Phys. B: At. Mol. Opt.
  Phys.} \bibinfo{volume}{49} (\bibinfo{year}{2016}) \bibinfo{pages}{145004--1
  -- 145004--10}. \DOIprefix\doi{10.1088/0953-4075/49/14/145004}.
\bibitem[{Barbosa et~al.(2015)Barbosa, Almeida, and Prudente}]{barbosa}
\bibinfo{author}{T.~N. Barbosa}, \bibinfo{author}{M.~M. Almeida},
  \bibinfo{author}{F.~V. Prudente}, \bibinfo{journal}{J. Phys. B: At., Mol.
  Opt. Phys.} \bibinfo{volume}{48} (\bibinfo{year}{2015})
  \bibinfo{pages}{055002--1 -- 055002--10}.
  \DOIprefix\doi{10.1088/0953-4075/48/5/055002}.
\bibitem[{Nascimento and Prudente(2016)}]{wallas_fred_educacional}
\bibinfo{author}{W.~S. Nascimento}, \bibinfo{author}{F.~V. Prudente},
  \bibinfo{journal}{Quim. Nova} \bibinfo{volume}{39} (\bibinfo{year}{2016})
  \bibinfo{pages}{757--764}. \DOIprefix\doi{10.5935/0100-4042.20160045}.
\bibitem[{Laguna and Sagar(2014)}]{uncertainties_of_the_confined_harmonic}
\bibinfo{author}{H.~G. Laguna}, \bibinfo{author}{R.~P. Sagar},
  \bibinfo{journal}{Ann. Phys. (Berlin, Ger.)} \bibinfo{volume}{526}
  (\bibinfo{year}{2014}) \bibinfo{pages}{555--566}.
  \DOIprefix\doi{10.1002/andp.201400156}.
\bibitem[{Bialynicki-Birula and Mycielski(1975)}]{relacaodeincertezainformacao}
\bibinfo{author}{I.~Bialynicki-Birula}, \bibinfo{author}{J.~Mycielski},
  \bibinfo{journal}{Commun. Math. Phys.} \bibinfo{volume}{44}
  (\bibinfo{year}{1975}) \bibinfo{pages}{129--132}.
  \DOIprefix\doi{10.1007/BF01608825}.
\bibitem[{Massen et~al.(2002)Massen, Moustakidis, and
  Panos}]{coparacaofermionboson}
\bibinfo{author}{S.~Massen}, \bibinfo{author}{C.~Moustakidis},
  \bibinfo{author}{C.~Panos}, \bibinfo{journal}{Phys. Lett. A}
  \bibinfo{volume}{299} (\bibinfo{year}{2002}) \bibinfo{pages}{131--136}.
  \DOIprefix\doi{10.1016/S0375-9601(02)00667-9}.
\bibitem[{Site(2014)}]{shannon_correlacao3}
\bibinfo{author}{L.~D. Site}, \bibinfo{journal}{Int. J. Quantum Chem.}
  \bibinfo{volume}{115} (\bibinfo{year}{2014}) \bibinfo{pages}{1396--1404}.
  \DOIprefix\doi{10.1002/qua.24823}.
\bibitem[{Gadre and Bendale(1985)}]{gadrequalidade}
\bibinfo{author}{S.~Gadre}, \bibinfo{author}{R.~Bendale},
  \bibinfo{journal}{Curr. Sci.} \bibinfo{volume}{54} (\bibinfo{year}{1985})
  \bibinfo{pages}{970--977}.
\bibitem[{Gadre et~al.(1985)Gadre, Sears, Chakravorty, and
  Bendale}]{gardre1985}
\bibinfo{author}{S.~R. Gadre}, \bibinfo{author}{S.~B. Sears},
  \bibinfo{author}{S.~J. Chakravorty}, \bibinfo{author}{R.~D. Bendale},
  \bibinfo{journal}{Phys. Rev. A} \bibinfo{volume}{32} (\bibinfo{year}{1985})
  \bibinfo{pages}{2602--2606}. \DOIprefix\doi{10.1103/PhysRevA.32.2602}.
\bibitem[{H{\^o} et~al.(1994)H{\^o}, Sagar, Pérez-Jord{\'a}, {Smith Jr.}, and
  Esquivel}]{estudo_numerico_qualidade_base}
\bibinfo{author}{M.~H{\^o}}, \bibinfo{author}{R.~P. Sagar},
  \bibinfo{author}{J.~M. Pérez-Jord{\'a}}, \bibinfo{author}{V.~H. {Smith
  Jr.}}, \bibinfo{author}{R.~O. Esquivel}, \bibinfo{journal}{Chem. Phys. Lett.}
  \bibinfo{volume}{219} (\bibinfo{year}{1994}) \bibinfo{pages}{15--20}.
  \DOIprefix\doi{10.1016/0009-2614(94)00029-8}.
\bibitem[{Guevara et~al.(2003)Guevara, Sagar, and
  Esquivel}]{shannon_correlacao}
\bibinfo{author}{N.~L. Guevara}, \bibinfo{author}{R.~P. Sagar},
  \bibinfo{author}{R.~O. Esquivel}, \bibinfo{journal}{Phys. Rev. A}
  \bibinfo{volume}{67} (\bibinfo{year}{2003}) \bibinfo{pages}{12507--1 --
  12507--6}. \DOIprefix\doi{10.1103/PhysRevA.67.012507}.
\bibitem[{Matta et~al.(2011)Matta, Sichinga, and
  Ayers}]{teoriadainformacaodesistemasquanticos}
\bibinfo{author}{C.~F. Matta}, \bibinfo{author}{M.~Sichinga},
  \bibinfo{author}{P.~W. Ayers}, \bibinfo{journal}{Chem. Phys. Lett.}
  \bibinfo{volume}{514} (\bibinfo{year}{2011}) \bibinfo{pages}{379--383}.
  \DOIprefix\doi{10.1016/j.cplett.2011.08.072}.
\bibitem[{Flores-Gallegos(2016)}]{aproximacao_dimencao_shannon}
\bibinfo{author}{N.~Flores-Gallegos}, \bibinfo{journal}{Chem. Phys. Lett.}
  \bibinfo{volume}{650} (\bibinfo{year}{2016}) \bibinfo{pages}{57--59}.
  \DOIprefix\doi{10.1016/j.cplett.2016.02.061}.
\bibitem[{Dehesa et~al.(2002)Dehesa, Mart{\'i}nez-Finkelshtein, and
  Sorokin}]{quantuminformacao}
\bibinfo{author}{J.~S. Dehesa}, \bibinfo{author}{A.~Mart{\'i}nez-Finkelshtein},
  \bibinfo{author}{V.~N. Sorokin}, \bibinfo{journal}{Phys. Rev. A}
  \bibinfo{volume}{66} (\bibinfo{year}{2002}) \bibinfo{pages}{062109--1 --
  062109--7}. \DOIprefix\doi{10.1103/PhysRevA.66.062109}.
\bibitem[{Orlowski(1997)}]{teoriadainformacaoesqueezingdeflutuacoes}
\bibinfo{author}{A.~Orlowski}, \bibinfo{journal}{Phys. Rev. A}
  \bibinfo{volume}{56} (\bibinfo{year}{1997}) \bibinfo{pages}{2545--2548}.
  \DOIprefix\doi{10.1103/PhysRevA.56.2545}.
\bibitem[{Aquilanti et~al.(1997)Aquilanti, Cavalli, and
  Coletti}]{aquilanti1997}
\bibinfo{author}{V.~Aquilanti}, \bibinfo{author}{S.~Cavalli},
  \bibinfo{author}{C.~Coletti}, \bibinfo{journal}{Chem. Phys.}
  \bibinfo{volume}{214} (\bibinfo{year}{1997}) \bibinfo{pages}{1--13}.
  \DOIprefix\doi{10.1016/S0301-0104(96)00310-2}.
\bibitem[{Sakurai and Napolitano(2011)}]{sakurai}
\bibinfo{author}{J.~J. Sakurai}, \bibinfo{author}{J.~Napolitano},
  \bibinfo{title}{Modern Quantum Mechanics},
  \bibinfo{publisher}{Addison-Wesley}, \bibinfo{address}{San Francisco},
  \bibinfo{year}{2011}.
\bibitem[{Luden{\~a}(1977)}]{ludena}
\bibinfo{author}{E.~V. Luden{\~a}}, \bibinfo{journal}{J. Chem. Phys.}
  \bibinfo{volume}{66} (\bibinfo{year}{1977}) \bibinfo{pages}{468--470}.
  \DOIprefix\doi{10.1063/1.433964}.
\bibitem[{Varshni(1997)}]{varshni}
\bibinfo{author}{Y.~P. Varshni}, \bibinfo{journal}{J. Phys. B: At., Mol. Opt.
  Phys.} \bibinfo{volume}{30} (\bibinfo{year}{1997})
  \bibinfo{pages}{L589--L593}. \DOIprefix\doi{10.1088/0953-4075/30/18/001}.
\bibitem[{Griffiths(1995)}]{griffiths}
\bibinfo{author}{D.~J. Griffiths}, \bibinfo{title}{Introduction to Quantum
  Mechanics}, \bibinfo{publisher}{Prentice Hall}, \bibinfo{address}{New
  Jersey}, \bibinfo{year}{1995}.

\end{thebibliography}

\end{document}